\documentclass[aps,prl,preprint]{revtex4}
\usepackage{graphicx}
\renewcommand{\r}{ {\bf{r}} }

\begin{document}
\title{Systematically Improvable Generalization of Self-Interaction Corrected Density Functional Theory}
\author{Benjamin G. Janesko}
\affiliation{Department of Chemistry \& Biochemistry, Texas Christian University, 2800 S. University Dr, Fort Worth, TX 7629, USA} 
\date{\today}

\begin{abstract} 
Perdew-Zunger self-interaction correction (PZSIC) reintroduces an exact
constraint to approximate density functional theory (DFT), but can
paradoxically degrade performance and is not systematically improvable.  We use
the Adiabatic Projection formalism to derive PZSIC in terms of a reference
system experiencing only electron self-interaction.  Generalization introduces
correlation into the reference system, systematically bridging from PZSIC to
exact wavefunction theory.  Minimal active spaces resolve the PZSIC paradox,
accurately treating near-equilibrium and strongly-correlated systems. 

\end{abstract}

\maketitle

Kohn-Sham density functional theory is the most widely-used electronic
structure approximation. KS-DFT
models a real system of interacting electrons in terms of a reference system of
noninteracting Fermions, corrected by mean-field (Hartree) electron repulsion
and a formally exact exchange-correlation (XC) spin-density functional.
Standard approximate XC functionals typically work well for properties such as
lattice constants and bond energies, but can fail dramatically for properties
such as band gaps, some reaction barriers, charge-transfer excited states, and
stretched bonds.\cite{Perdew2009,Zope2019} Failures typically involve (1)
nearly-one-electron regions where approximate XC functionals do not cancel the
spurious Hartree self-interaction, and (2) strongly correlated regions where the
exact reference system wavefunction substantially differs from the exact real
system wavefunction.\cite{Janesko2021}
 
The Perdew-Zunger self-interaction correction\cite{Perdew1981} addresses
failure (1) by removing self-interaction error in an
orbital-by-orbital fashion, 
\begin{eqnarray}
E_{XC}^{SICSL}[\rho_\uparrow,\rho_\downarrow] &=& E_{XC}^{SL}[\rho_\uparrow,\rho_\downarrow]-\sum_{i,\sigma}^{occ}U[\rho_{i\sigma}]+E_{XC}^{SL}[\rho_{i\sigma},0]. 
\end{eqnarray}
$\rho_\sigma(\r)=\sum_{i=1}^{N\sigma}|\psi_{i\sigma}(\r)|^2$ is the probability
density of the $N_\sigma$ $\sigma=\uparrow,\downarrow$-spin electrons,
$E_{XC}^{SL}$ is a semilocal\cite{Burke1998} approximation to the exact XC
functional, self-Hartree energy $U[\rho_{i\sigma}]=(1/2)\int d^3\r_1\int d^3\r_2\rho_{i\sigma}(\r_1)|\r_1-\r_2|^{-1}\rho_{i\sigma}(\r_2)$,  and the reference system's single Slater determinant wavefunction is
composed of spin-orbitals $\{\psi_{i\sigma}\}$. Orbital densities
$\rho_{i\sigma}=|\phi_{i\sigma}|^2$ are constructed from localized orbitals
$\{\phi_{i\sigma}\}$ defined as a unitary transform of $\{\psi_{i\sigma}\}$.
While optimizing this transform can be expensive,\cite{Pederson1984,Vydrov2004}
complex\cite{Jonsson2011} or Fermi-L\"owdin\cite{Pederson2014}  orbitals have 
made SIC increasingly practical.\cite{Sharkas2020}
 
This has led to a renewed focus on the "paradox" of self-interaction.
Semilocal functionals' self-interaction simulates some opposite-spin
correlation, thus PZSIC can degrade performance for systems near equilibrium
and especially for strongly correlated systems.\cite{Zope2019} This exemplifies
the broader "zero-sum" tradeoffs of standard XC approximations.\cite{Janesko2021}
One road beyond this tradeoff is to reintroduce part of the electron-electron
interaction into the reference system.\cite{Savin1988,Fromager2007} 

Here we address the "paradox" of self-interaction by first
explicitly deriving the PZSIC {\em{ansatz}} to address failure (1), then generalizing 
to address failure (2).  The derivation uses the Adiabatic
Projection formalism.\cite{Janesko2022a} Consider $N$ electrons in
external potential $v_{ext}(\r)$.  Define $\hat{T}$ and  $\hat{V}_{ee}$ as the
real system's kinetic and electron-electron repulsion operators. Define
$\hat{P}_{i\sigma}=\left|\phi_{i\sigma}\phi_{i\sigma}\right>\left<\phi_{i\sigma}\phi_{i\sigma}\right|$
as a two-electron projection onto localized orbital $\phi_{i\sigma}$. (This denotes the direct product: 
$\left<\r_1,\r_2|\phi_{i\sigma}\phi_{i\sigma}\right>=\phi_{i\sigma}(\r_1)\phi_{i\sigma}(\r_2)$.)
Define a projected operator
$\hat{V}^{P}_{ee}=\sum_{i\sigma}\hat{P}_{i\sigma}\hat{V}_{ee}\hat{P}_{i\sigma}$.
Introduce an adiabatic connection parameter $\lambda$ and a generalized energy functional\cite{Levy1985a}
\begin{eqnarray}
\label{eq:gdf}
E[\lambda,\{\hat{P}_{i\sigma}\},\rho] &=& \int d^3\r \rho(\r)v_{ext}(\r)+\min_{\Psi\to\rho}\left<\Psi\right|\hat{T}+\hat{V}^{P}_{ee}+\lambda(\hat{V}_{ee}-\hat{V}^P_{ee})\left|\Psi\right>
\end{eqnarray}
Here $\lambda=1$ is the real system and $\lambda=0$ is a reference system
experiencing only the projected electron-electron repulsion operator. Define
$\Psi[\lambda,\{\hat{P}_{i\sigma}\},\rho]$ as the minimizing wavefunction of
Eq. \ref{eq:gdf}.  The
reference system experiences only electron self-interaction, thus its
electron-electron interaction energy is zero and its ground-state wavefunction
$\Phi[\lambda=0,\{\hat{P}_{i\sigma}\},\rho]$ is 
a single Slater determinant composed of the spin-orbitals
$\{\psi_{i\sigma}\}$ introduced above. Explicitly, 
\begin{eqnarray}
E_{ee}[\lambda=0,\{\hat{P}_{i\sigma}\},\rho]&=& \left<\Phi[\lambda=0,\{\hat{P}_{i\sigma}\},\rho]|\sum_{i\sigma} \hat{P}_{i\sigma}\hat{V}_{ee}\hat{P}_{i\sigma}|\Phi[\lambda=0,\{\hat{P}_{i\sigma}\},\rho]\right> \\ 
&=& \sum_{i\sigma jk} c^*_{ji}c^*_{ki}U[\rho_{i\sigma}]\left(c_{ij}c_{ik}-c_{ik}c_{ij}\right) \ = \  \sum_{i\sigma} U[\rho_{i\sigma}]+E_X^{ex}[\rho_{i\sigma}]  \nonumber 
\end{eqnarray}
Here  $c_{ij}=\left<\phi_{i\sigma}|\psi_{j\sigma'}\right>$ ensuring 
$\left(c_{ij}c_{ik}-c_{ik}c_{ij}\right)$ is zero. Self-Hartree energy
$U[\rho_{i\sigma}]=
\left<\phi_{i\sigma}\phi_{i\sigma}|\hat{V}_{ee}|\phi_{i\sigma}\phi_{i\sigma}\right>$ was
introduced above. Self-exchange
$E_{X}^{ex}[\rho_{i\sigma}]=-U[\rho_{i\sigma}]$. In PZSIC, localized orbitals
$\{\phi_{i\sigma}\}$ are a unitary transform of $\{\psi_{i\sigma}\}$ ensuring
$\sum_jc^*_{ji}c_{ij}=1$.
The Hohenberg-Kohn theorems ensure that the real system's ground-state density
and energy can be obtained by minimizing the energy of the reference system,
corrected by some density functional
\begin{eqnarray}
\label{eq:HXC}
E_{HXC}[\{\hat{P}_{i\sigma}\},\rho] &=&\int_0^1d\lambda\left<\Psi[\lambda,\{\hat{P}_{i\sigma}\},\rho]\right|\hat{V}_{ee}-\hat{V}^P_{ee}\left|\Psi[\lambda,\{\hat{P}_{i\sigma}\},\rho]\right> \\ 
&=& U[\rho]-\left(\sum_{i\sigma}U[\rho_{i\sigma}]\right)+E_{XC}[\{\hat{P}_{i\sigma}\},\rho] \nonumber 
\end{eqnarray}
Ref. \cite{Janesko2022a} gives details of this adiabatic connetion. The
$U[\rho]$ and $U[\rho_{i\sigma}]$ terms in eq \ref{eq:HXC} are respectively from the
$\lambda=1$ and $\lambda=0$ limits of the adiabatic connection. 
Approximating the projected XC functional as $E_{XC}[\{\hat{P}_{i\sigma}\},\rho]
\simeq
E_{XC}^{SL}[\rho_\uparrow,\rho_\downarrow]-\sum_{i\sigma}E_{XC}^{SL}[\rho_{i\sigma},0]$
recovers the PZSIC. 

\begin{figure}
\includegraphics[width=0.5\textwidth]{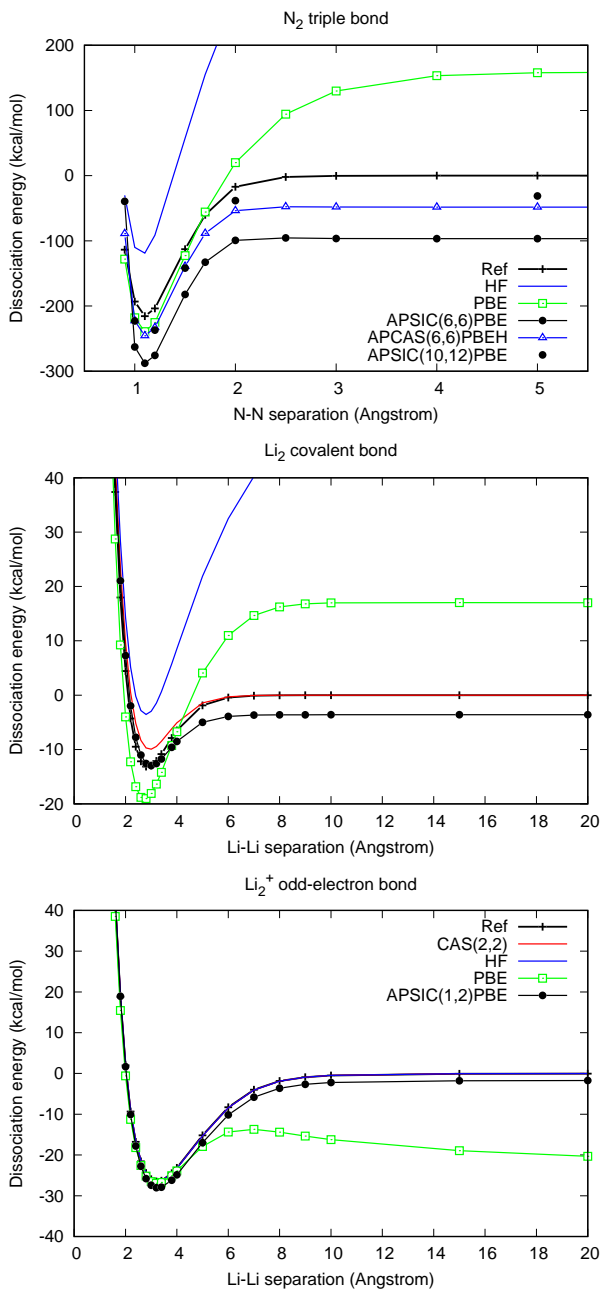}
\caption{\label{fig:Li} Dissociation curves of Li$_2^+$ (top), spin-restricted singlet Li$_2$ (middle), and spin-restricted singlet N$_2$ (bottom), dissociation energies relative to separate atoms.}
\end{figure}

\begin{figure}
\includegraphics[width=0.65\textwidth]{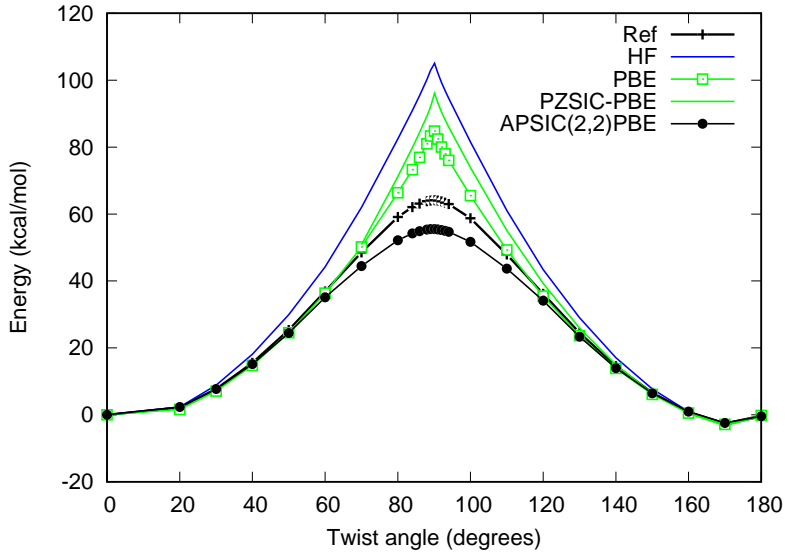}
\caption{\label{fig:twist} Torsional potential energy surface for 2-butene
double bond rotation.} \end{figure}

With this derivation in hand, generalization of PZSIC to treat failure (2) merely
requires generalizing the projections from single orbitals to active spaces.
We illustrate a closed-shell $N_\sigma$-electron system such
as  stretched singlet Li$_2$
(Figure \ref{fig:Li}) or twisted 2-butene (Figure \ref{fig:twist}), with a single active
space of two electrons in two orbitals (four spin-orbitals)
$\psi_{h\sigma}$ and $\psi_{l\sigma}$. Choose $\{\phi_{i\sigma}\}$ as a unitary
transform of the $N_\sigma-1$ remaining "core" orbitals.  Define a
projection onto the active space and a projected
electron-electron repulsion operator, 
\begin{eqnarray}
\label{eq:p22} 
\hat{P}_{22} &=& \sum_{a,b=h,l} \sum_{\sigma,\sigma'} \left|\psi_{a\sigma}\psi_{b\sigma'}\right>\left<\psi_{a\sigma}\psi_{b\sigma'}\right|,  \\ 
\label{eq:v22}
\hat{V}^{P22}_{ee} &=&\hat{P}_{22}\hat{V}_{ee}\hat{P}_{22} + \sum_\sigma \sum_{i=1}^{N\sigma-1}\hat{P}_{i\sigma}\hat{V}_{ee}\hat{P}_{i\sigma}. 
\end{eqnarray}
The reference system's electron-electron interaction energy is now
nonzero, and its exact ground state $\Psi^{CAS}[\lambda=0,\rho]$ is a
multireference complete active space CAS(2,2) wavefunction capable of giving
noninteger active orbital occupancies $n_{h\sigma},n_{l\sigma}$. The real
system's electron-electron interaction energy is modeled as 
\begin{eqnarray}
\label{eq:cas22} 
E_{ee} &=&
\left<\Psi^{CAS}[\lambda=0,\rho]|\hat{V}_{ee}|\Psi^{CAS}[\lambda=0,\rho]\right>
+ (E_{XC}^{SL}[\rho]-E_{XC}^{SL}[\rho_a])  \\ 
&& -(E_X^{ex}[\rho]-E_X^{ex}[\rho_a])   -\sum_\sigma\sum_{i\sigma=1}^{N\sigma-1}\left(U[\rho_{i\sigma}] + E_{XC}^{SL}[\rho_{i\sigma},0]  \right) \nonumber 
\end{eqnarray}
Here $\rho_a=n_{h}|\psi_{h}|^2+n_{l}|\psi_{l}|^2$ is the electron density
projected onto the active space.  The first expectation value in Eq
\ref{eq:cas22} includes the Hartree energy $U[\rho]$, multireference exchange and
correlation in the active space, and exact exchange outside the active space.  The
other terms add semilocal XC
outside the active space, subtract off exact exchange outside the active space,
and apply the SIC to the remaining core orbitals. The approach is systematically improvable in the way that
correlated wavefunction theory is systematically improvable, recovering full
configuration interaction (FCI) in the limit where the active space projector
includes all occupied and unoccupied orbitals. It explicitly avoids
"double-counting" of exchange-correlation, just as PZSIC avoids "double-counting" of
exchange-correlation.  We use "APSIC($n,m$)SL" to
denote generalized self-interaction correction with a single active space of
$n$ electrons and $m$ orbitals, using semilocal functional "SL".
"APCAS($n,m$)SL" denotes omitting the core-orbital SIC (the sums over $i\sigma$
in Eq.
\ref{eq:cas22}).

Figure \ref{fig:Li} shows APSIC($n,2$)PBE dissociation curves of Li$_2^+$
($n=1$), a system where PZSIC fixes failure (1), and spin-restricted singlet
Li$_2$ ($n=2$), a system where PZSIC worsens failure (2).  Calculations use the
def2-TZVP basis set,\cite{Weigend2005} the PySCF package,\cite{Sun2017,Sun2020}
the Perdew-Burke-Ernzerhof (PBE) semilocal XC functional,\cite{Perdew1996}
CAS($n,m$)SCF canonical orbitals, and Edmiston-Ruedenberg localized core
orbitals\cite{Edmiston1963} to compute the core-orbital SIC. Dissociation
energies are relative to separated atoms, evaluated with $n=0$. "Ref" denotes
N-electron valence perturbation theory with the CAS($n,m$)SCF
reference.\cite{Angeli2001} 
Code to perform all calculations is freely
available at github.org/bjanesko/apsic.  

The generalized SIC fixes both failures of standard DFT. APSIC($1,2$)PBE
recovers PZSIC for Li$_2^+$, fixing the overbinding of PBE (failure (1)).
APSIC($2,2)$PBE gives a near-exact dissociation limit for strongly correlated
spin-restricted Li$_2$ (failure (2)), and is close to the reference curve near
equilibrium.  

The 3.6 kcal/mol APSIC(2,2)PBE overbinding of dissociated Li$_2$
arises not from "double-counting", but from PBE overbinding outside of the
active space.  PBE (HF) calculations on stretched Li$_2$ give an XC
contribution to the bond energy $E_{XC}(Li_2)-2E_{XC}(Li)$ of +13.6 (+73.4)
kcal/mol, with +17.2 (+73.4) kcal/mol from the active space, leaving a
difference of 3.6 (0.0) kcal/mol.  PBE "overbinds" the dissociation limit
relative to HF theory, and not all of that overbinding is within the active
space. Correcting the active space with wavefunction correlation leaves a residual
overbinding. The bottom panel of Figure \ref{fig:Li} shows similar effects for
the stretched singlet symmetry-restricted N$_2$ triple bond. APSIC(6,6)PBE
overbinding can be fixed by increasing the active space, or with the
APCAS(6,6)PBEH "half-and-half" global hybrid.\cite{Becke1993} Pair density
functionals address related effects in other wavefunction-in-DFT
methods.\cite{Sharma2021,Perdew1995}

Figure \ref{fig:twist} shows the APSIC(2,2)PBE torsional potential energy
surface for 2-butene double bond rotation. Calculations use the 6-31G(d) basis
set and spin-unrestricted UB3LYP/6-31G* geometries optimized in Gaussian
16.\cite{g16} Hartree-Fock, PBE, and PZSIC-PBE calculations give an unphysical
cusp due to failure (2). APSIC(2,2)PBE is instead quite close to the reference. 

These results suggest a number of avenues for future work. 
(1)  The Adiabatic Projection derivation admits many choices of projection
$\hat{P}_{i\sigma}$, including the complex-valued\cite{Lehtola2016} or
Fermi-L\"owdin\cite{Pederson2014} localized orbitals of modern SIC.  This
derivation works with those methods. Moreover, there is no requirement to
choose the localized orbitals as a unitary transform of the canonical orbitals.
Other choices, such as predefined atomic orbitals, may provide formal
connections to methods such as DFT+U.\cite{Kulik2015} 
(2) The present approximation does not include active-virtual correlation.
APSIC(2,2)PBE calculations on a closed-shell two-electron system recover
CAS(2,2), not FCI. (Of course, increasing the active space is guaranteed to
converge to FCI.) Active-virtual correlations could be added using existing
wavefunction\cite{Chatterjee2016} or wavefunction-in-DFT\cite{Graefenstein2005}
approaches.
(3) Introducing multiple orthogonal active spaces provides connections to other
correlated wavefunction model chemistries.  For example, while treating an
$N$-electron system with $N$ mutually orthogonal {\em{one}}-orbital spaces
recovers PZSIC, $N$ mutually orthogonal {\em{two}}-orbital spaces could recover
a "DFT-corrected" version of perfect pairing / antisymmetrized geminal product
approximations.\cite{Beran2005,Pernal2013} Localization of the active space
orbitals provides other connections to linear-scaling localized-orbital {\em{ab
initio}} methods.\cite{Scuseria1999} Overall, the results motivate further
application of the Adiabatic Projection formalism to computationally tractable
model chemistries beyond standard DFT.

\bibliographystyle{aip}

\end{document}